\documentclass[11pt,letterpaper]{article}
\usepackage[T1]{fontenc}
\usepackage[utf8]{inputenc}
\usepackage[margin=1in]{geometry}
\usepackage{mathpazo}
\usepackage{microtype}
\usepackage{booktabs,tabularx,array}
\usepackage{graphicx}
\graphicspath{{figures/}}
\usepackage[super,sort&compress]{natbib}
\usepackage{xurl}
\usepackage[hidelinks]{hyperref}
\title{Reproducibility in the Age of Agentic AI:\\ Context Engineering at the Timescale of a Codebase}
\author{Lorena A. Barba\\[0.4em]
\small The George Washington University, Washington, D.C.}
\date{}
\begin{document}
\maketitle
\begin{abstract}
Reproducible research practices are context engineering for AI coding agents. I argue that agents lower the cost of maintaining tests, commit histories, repository structure, instructions, and decision records while making their benefits immediate. Researchers remain responsible for verifying these artifacts and the scientific judgments they encode.
\end{abstract}

Since the start of my career as a principal investigator (PI), I have advocated for reproducible research practices, the upshot of the toils of my own PhD studies.\cite{barba2016} Version control, open code development and public release, sharing of research artifacts like figures and data, and posting preprints were all part of my Reproducibility PI Manifesto more than a decade ago.\cite{barba2012} My group added habits of code testing, clean repository layouts, and commit messages that explain \emph{why} and not just \emph{what}, and I encouraged them in every student cohort, course, and collaboration. Yet despite copious writing and speaking about reproducibility, by myself and many others, uptake has been modest. Friction remains a powerful argument: each practice costs time and effort in every session but pays back only over many, and graduate students live under deadline pressure that punishes long-horizon investment. The research community has accepted this as the cost of doing science with people who are not trained or paid to be software engineers. I admit to becoming disenchanted over the years, even a bit cynical. It came as a revelation, then, to find that the calculus has changed and that generative AI is spurring new interest in reproducibility. In this paper, I put forward an updated view of computational reproducibility for the era of agent-driven code development.

\section{Context Is Everything for AI Agents}

While preparing a class for my recent course on Generative AI for Engineering Research,\footnote{\url{https://barbagroup.github.io/mae6291-genai/}} I was working through the exercise of how to dissect an existing agentic system and explain its parts through the vocabulary of \textbf{context engineering}: the practice of deciding what information reaches a model, in what form, and when, given that the model acts only on what is in its limited input window, its context. The exercise was meant to be diagnostic: take a working automation built on a large language model (LLM), name its context-engineering decisions, identify where the design is principled and where it is incidental. What emerged was a bit of a surprise. \emph{Every} significant design choice in the system could be expressed in terms of context. Borrowing the terms psychologists use for human memory, the static files the agent reads at startup are its always-on \emph{semantic memory}: durable facts and rules, present in every session. The structured outputs it produces are its \emph{episodic} record: an account of what happened, and when, that a later session can consult. The pre-processing applied before content reaches the model is \emph{attention-budget management}: a model attends well to only a limited amount of context, so what is filtered or condensed beforehand leaves more of that budget for the task. The narrowing of the agent's tool surface, the set of actions it is allowed to take, is \emph{scope control}. After working through the system components like this, there was nothing left over. A working agentic system is a collection of context-engineering decisions: stacked in layers, made at different times, often by different hands, but reducible to that single discipline.

Once you can read a system as an implementation of context engineering, you can flip the question from diagnosis to design: how might you build a system so that an agent can be maximally productive in it? Asking that question seriously leads into new territory, with both serious concerns and clear opportunities.

\section{Good Software Practices Are Now Context Engineering}

The next surprise was that the answer to ``how do I make this system agent-friendly?'' was largely the answer to ``how do I make this system reproducible?'' An agent instruction file, a test suite, a clean commit history, a well-organized repository, and a decision record: each maps onto a category of agent context (Table~\ref{tab:artifacts}). The mapping is strict, not a loose analogy. These artifacts are \textbf{context engineering at the timescale of a codebase} rather than the timescale of a conversation. Good software practices have thus acquired a new and entirely mechanical reason for existing: agents need them to work productively and correctly.

\subsection{Catalog of Agent-Facing Artifacts}

\begin{figure}[htbp]
\centering
\includegraphics[width=\textwidth]{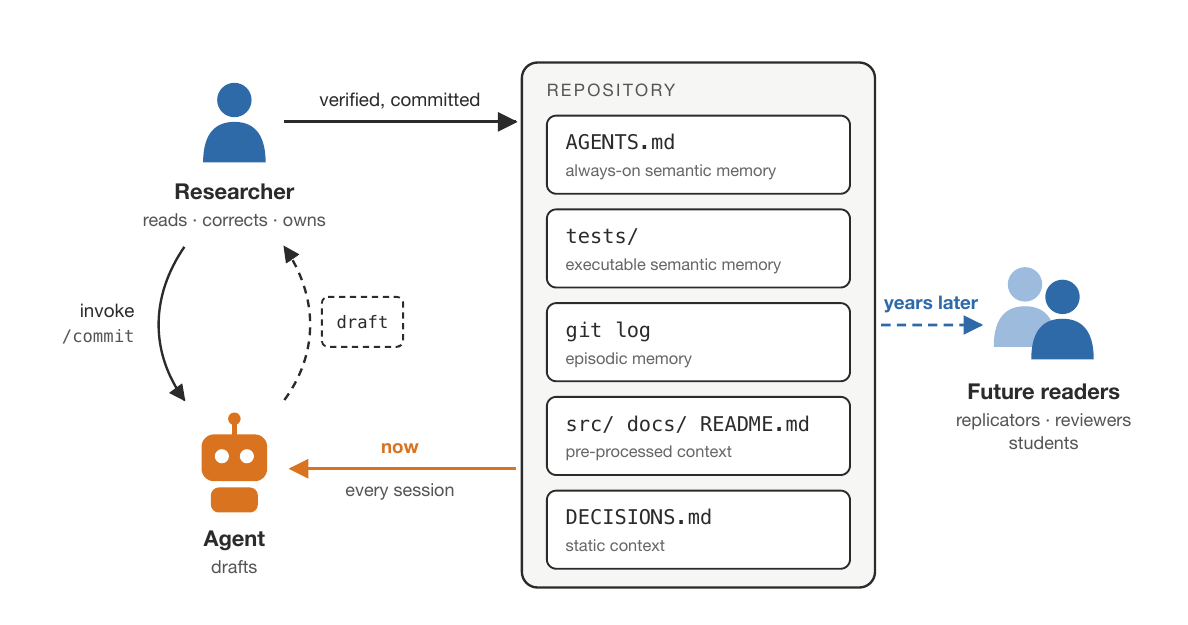}
\caption{The draft--verify loop. The researcher invokes the agent, which drafts an artifact; the researcher reads, corrects, and takes ownership of the draft before committing it to the repository. The committed artifacts, labeled with their context categories from Table~\ref{tab:artifacts}, then serve two readers: the agent, at the start of every session, and future human readers, years later.}
\label{fig:loop}
\end{figure}

The five artifacts in Table~\ref{tab:artifacts} predate LLMs by years or decades, and each was advocated, with limited success, throughout the era when research code was read only by humans. Two things are new. First, the cost of producing them has collapsed, because an agent drafts them on request. Your role shifts from author to \textbf{owner and verifier}: the agent drafts the test, the message, the record, and you read it, correct it, and take responsibility for it (Figure~\ref{fig:loop}). The artifact is no less yours for having been drafted by a machine, just as a number produced by a solver you wrote is still your result. You do not need to type the code to own the judgments involved. Second, the artifacts have acquired a second reader, the agent itself, which consults them at the start of and throughout every session. An artifact that used to pay back slowly over the life of a project now also pays back immediately, in the quality of the next interaction. Peers benefit too: supported by their own agents, they can more easily reproduce and build upon openly shared work.

\begin{table}[htbp]
\caption{Agent-facing artifacts in a research software repository: the context category each belongs to, its two functions, and how it is now produced.}
\label{tab:artifacts}
\centering
\footnotesize
\setlength{\tabcolsep}{4pt}
\renewcommand{\arraystretch}{1.15}
\begin{tabularx}{\textwidth}{@{}*{5}{>{\raggedright\arraybackslash}X}@{}}
\toprule
\textbf{Artifact} & \textbf{Context category} & \textbf{Reproducibility function} & \textbf{Agent-friendliness function} & \textbf{Now produced by} \\
\midrule
Agent instruction file: \texttt{AGENTS.md / CLAUDE.md} & Always-on semantic memory & Conventions of record: how the project is built and run & Startup context every session; no re-explanation & Agent drafts from the repo; you add the tacit parts \\
Test suite (characterization tests) & Executable semantic memory & Behavior made verifiable, not asserted & Self-check signal the agent runs on its own work & Agent drafts tests for trusted functions; you confirm intent \\
Commit history (Conventional Commits) & Episodic memory & Provenance of every change & Recent trajectory reconstructed from \texttt{git log} & Agent drafts the message from the diff; you edit and own it \\
Repository structure & Pre-processed context & A layout that can be inherited, not abandoned & Minimal exploration; attention spent on the task & Agent scaffolds from a prompt; you adjust \\
Decision record (DECISIONS.md) & Static context & The reasoning that links intent to implementation & Settled questions are not relitigated & Agent drafts the entry; you supply the reasoning \\
\bottomrule
\end{tabularx}
\end{table}

\textbf{The agent instruction file} (\texttt{AGENTS.md} or \texttt{CLAUDE.md}) is your project's always-on semantic memory: the document an agent reads at the start of every session, encoding what the project is, how it is built and run, the conventions it follows, and the boundaries it must respect. A vague file (``a helpful coding assistant'') yields vague behavior. A GitHub analysis of more than 2,500 such files in public repositories found that effective ones are specific, and that the most useful convention is a three-tier boundary, \emph{always do, ask first, never do}, which turns the file from documentation into operative instruction.\cite{nigh2025} You do not write it from scratch: the agent reads the repository and drafts it, and you supply the tacit knowledge that is invisible in the code, such as which data files are sacred or why a particular numerical convention is firmly adopted.

\textbf{The test suite} is executable semantic memory: a specification of expected behavior in a form that can be run and checked. For research code the most valuable pattern is the \textbf{characterization test}, which captures the current, trusted behavior of a function without trying to prove it theoretically correct.\footnote{The term originates with Michael Feathers, \emph{Working Effectively with Legacy Code} (2004).\cite{feathers2004} See the review on the Better Scientific Software blog.\cite{bartlett2019}} Thirty characterization tests of the thirty functions whose correctness carries your scientific claims are worth more than three hundred tests of trivial accessors. This is where an agent's help can be crucial, because writing the first test was historically the highest-friction practice of all: you name a function you already trust, the agent drafts tests that pin its behavior, and you check that the behavior it captured is the behavior you intended. The suite then turns every later change, yours or the agent's, into a checkable event, closing the loop between generation and verification.

\textbf{The commit history} is episodic memory: an ordered account of what changed, when, and why. The \emph{Conventional Commits} convention prefixes each message with a type---\texttt{feat, fix, refactor, docs, test, chore}---followed by an imperative summary and, where it matters, a body that explains the reasoning rather than restating the diff.\footnote{\url{https://www.conventionalcommits.org/}} The distance between ``update solver'' and ``fix(solver): correct sign of the Coriolis term in the v-momentum equation,'' followed by two lines on how the error survived the rotating-frame test and which published case finally exposed it, is the distance between a log that is noise and one that saves a successor two weeks of bisection eighteen months later. The agent removes the tedium that historically defeated this discipline: it reads the staged \texttt{diff} faster than you could describe it and drafts a message you edit and own. In return, a later agent session reconstructs the project's recent trajectory from the log in a few hundred tokens, rather than re-deriving it.

\textbf{The repository structure} is pre-processed context: the physical layout, which an agent reads simply by listing the directory tree. A repository ready for both human and machine readers has a predictable skeleton---a \texttt{README}, an instruction file, a license, a changelog, a decision record, and a dependency specification at the root; source under \texttt{src/}, tests under \texttt{tests/}, runnable examples under \texttt{examples/}, documentation under \texttt{docs/}. The working test is whether a new lab member, or an agent on its first session, can orient in under a minute by reading only the root. This is the substance of what has been called \textbf{AI-friendly codebase design}: the shape of the repository is itself a context interface, and its quality sets how much of every later interaction is spent on the task rather than on exploration.\cite{boeckeler2026} An agent will scaffold the structure from a single prompt; your contribution is to fit it to the realities of the project.

\textbf{The decision record} (\texttt{DECISIONS.md}) is static context: a running log of consequential choices and the reasoning behind each, in the format long used for \textbf{Architecture Decision Records} (ADRs)---context, decision, alternatives considered, consequences.\footnote{Michael Nygard introduced the Architecture Decision Record in a 2011 post; the community-maintained documentation is at \url{https://adr.github.io/}.} It is the rarest of these artifacts in research software repositories and the most valuable for agentic work, because it preserves exactly what neither the code nor the tests can show: the reasoning that connected a goal to the particular implementation chosen to meet it. The example below shows the form:

\begin{quote}
\begin{minipage}{\linewidth}
\setlength{\parskip}{0.5em}
\small
\textbf{2026-03-14: Use compressed sparse row (CSR) format throughout the linear solver}

\textbf{Context:} Memory became the binding constraint at grids above 512\textsuperscript{3}. Dense storage would exceed available RAM.

\textbf{Decision:} Use \texttt{scipy.sparse} CSR matrices for all stiffness and mass matrices. Do not convert to dense for debugging prints.

\textbf{Alternatives considered:} coordinate (COO) format (simpler, but slower matrix-vector products); dense (rejected for memory reasons); JAX sparse (rejected for dependency complexity).

\textbf{Consequences:} All solver code must tolerate sparse inputs. Visualization utilities that currently assume dense arrays will need \texttt{.toarray()} calls, documented in \texttt{CLAUDE.md}.

\end{minipage}
\end{quote}

Without such a record, returning to a convention months later---or meeting it for the first time, as an agent does---you cannot tell whether it was deliberate or incidental, and you will either keep it without understanding or discard it and walk straight back into the problem it was chosen to avoid. With the record, the decision becomes essential context that any reader, human or machine, recovers in seconds.

\subsection{Injecting Rigor: the Invocation Layer}

A reasonable objection emerges here. If an agent session runs for hours across dozens of turns, must you document every prompt to preserve a record of how the work was done? In my opinion, that target is both unmanageable and mistaken. The transcript of a long, exploratory session is not the artifact worth keeping; much of it could be dead ends, false starts, and reformulations, the very material you would never transcribe from your own thinking. What is worth keeping is what the catalog already identifies: the decisions, the verified behavior, the rationale, written to project files. \textbf{Provenance, in agentic research, lives in the maintained artifacts, not in the conversation that produced them.}

The practical question is how those artifacts stay current without your stopping the work to write them. The answer is a new stratum of the scientific programmer's literacy: controlled invocation. Coding agents now expose \textbf{slash commands} and \textbf{skills}, named and reusable operations that you trigger deliberately and that write to project files in a constrained, repeatable way. A \texttt{/commit} command drafts a Conventional Commits message from the staged diff; a decision-record skill appends a correctly formatted entry from a one-line statement of the choice and its rationale; a changelog skill updates the human-visible history at a release boundary. Each invocation yields a small, well-formed artifact as a byproduct of the work rather than as an interruption to it, and because the operation is named and reusable, it is applied consistently instead of only when you happen to remember. Learning the invocation layer---which commands and skills to define, and when to call them---is becoming as much a part of scientific programming as learning the shell once was. The catalog names the artifacts required for reproducibility; the invocation layer is how you produce them without the friction that used to stand in the way.

\subsection{Reproducibility Now Pays Off Immediately}

The practices I have been advocating for years, recognized by many as the right ones, failed to spread because the per-session cost was high and the long-horizon return was diffuse. The argument for reproducibility used to be: pay a cost today to benefit your future self, your peers, and your discipline. This is true and noble and also, demonstrably, was insufficient to motivate behavior change at scale. A new argument today is: pay almost no cost, because an agent will draft the artifact for you, and benefit \emph{both} your future self \emph{and} your present-day productivity, because the same agent uses the artifact every time you work. The practices themselves have not changed, nor have the failures they prevent: results that cannot be reproduced, codebases that cannot be modified, scientific claims that cannot be defended six months after publication. What has shifted is the bargain on offer to a working researcher, from ``do something hard and wait years for the payoff'' to ``let an agent help you make something easy and notice the payoff right away.'' If the reproducibility movement has been waiting for a tailwind, this could be one.

\section{A Convergence, Not a Coincidence}

The mapping in the previous section invites a skeptical reading: that it is a clever relabeling, reproducibility dressed in fashionable new vocabulary. A strong answer to that skepticism comes from people who were not thinking about reproducibility at all.

In the GitHub analysis introduced above, the authors set out to answer a purely instrumental question: across thousands of real repositories, what distinguishes an agent instruction file that makes a coding agent productive from one that does not? Examining more than 2,500 files, they found that the effective ones converge on six areas: commands, testing, project structure, code style, git workflow, and boundaries. Nothing in the study was motivated by scientific integrity, transparency, or the replication crisis; the goal was throughput, getting useful work out of an agent. These six areas are the same practices named with the artifact catalog, now seen from the agent's side of the table.

Read that list again with a reproducibility eye. Commands are the runnable instructions for building and executing the project, the ``how to run it'' that every reproducible-research checklist demands. Testing is verification, whose absence is the most commonly cited weakness of research software. Project structure is the organized, navigable layout that lets code be audited and inherited rather than abandoned. Git workflow is version control, the foundational reproducibility practice and the second pledge of my old reproducibility manifesto. Four of the six marks of an agent-ready repository are, almost item for item, practices the open-science movement has advocated for more than a decade.

The remaining two are softer matches. Code style maps to readability, a value reproducibility shares but has never placed at its center; it is closer to maintainability than to replication. Boundaries---the explicit list of what an agent must never touch---have the least clear analog, motivated as they are by the prevention of destructive automated mistakes. Even here the overlap is not nil: protecting raw data from being overwritten and separating sources from derived products is ordinary data-management hygiene, now made explicit because a new kind of reader can act on the files. Four clean matches and two partial ones: a striking convergence since no one engineered it.

What does it mean that two communities with unrelated motivations arrived at nearly the same result? The temptation is to call it coincidence; it is not. Both are solving the same underlying problem: how to make a codebase usable by a reader who arrives without the author's accumulated context. Reproducibility has always served such a reader: the replicator who was not in the lab, the reviewer who never met the author, the student who inherits the code two years after its author has gone. An AI agent is an extreme instance of that reader: it arrives with no shared history at the start of every session, and it arrives today rather than in two years. Practices that externalize an author's tacit knowledge into durable, legible artifacts serve all of these readers at once, whether the stranger consulting it is human or machine. In this sense reproducibility and agent-friendliness are the same thing: context engineering for a reader who lacks the author's context.

A third community has reached the same point from a different direction, motivated by caution. A recent vision statement from a research-software-engineering (RSE) workshop in Edinburgh, \emph{Research Software in an Age of AI-Assisted Development: Reflections from Edinburgh},\cite{katz2026} was prompted by concerns about the widening gap between a researcher's intent and an agent's implementation. It argues that the artifacts we expect for reproducibility must expand to capture not only code and data but the reasoning and the decisions that shaped them. Arriving from worry rather than enthusiasm, it reaches the same conclusion as the GitHub analysis: the durable record of how and why a system was built is what matters, and it has to be written down. When a movement built on advocacy, a study built on throughput, and a community animated by caution independently point to the same small set of artifacts, those artifacts are no longer one professor's preference or one company's house style. They are closer to a stable equilibrium: the configuration a research codebase settles into once it must answer to readers who do not share its author's mind.

\section{The Limits of the Optimistic View}

Readers would be right to distrust an argument that only collects reasons for optimism. The Edinburgh statement is the considered, collective worry of a community that has spent years on these questions. Several of its concerns are real, and the case I have made does not dissolve them.

The sharpest objection aims at the heart of the argument. If an agent drafts the test, the commit message, and the decision record, what stops a researcher from accepting all three without understanding any of them? The same tools that let a careful person externalize their context let a careless one outsource their judgment, and at scale the second pattern threatens the apprenticeship by which researchers once learned to verify and maintain code. This is the most serious risk, and I will not claim that any artifact prevents it. My claim is narrower: the artifacts are verification instruments, and verification is the one task in this workflow that cannot be delegated to an agent. A test you do not read is worthless; a decision record you did not reason through records nothing you know. The artifacts do not contribute understanding; they make understanding what you are responsible for (rather than the mechanics). The tool cannot supply the discipline, it only lower the cost of exercising it.

A second concern cuts at the word reproducibility itself. AI generation is probabilistic; models change underneath you; the same prompt on the same code may produce different output next month. The artifacts in the catalog don't fix that, but the argument operates at a different layer. The claim is not that the path by which code came to exist is reproducible; it is that the code, once it exists, is legible, tested, and accounted for. A characterization test does not care whether the function it pins was written by a human, by an agent, or by an agent on a model since deprecated. It asks only whether the behavior still holds. When a non-deterministic process produces a regression, the test suite is what tells you. The durable record, not the probabilistic act that produced it, is the part that science has always asked to inspect.

A third worry is that productivity is the wrong thing to celebrate in a research culture that already rewards volume over rigor, and that making good practices cheap will only let people produce mediocre work faster. Lowering the cost of quality is not a fix for incentives. Removing the excuse not to write a test does not make a hiring committee value the test, and an institution that counts papers will go on counting papers whether or not the underlying code is sound. My argument removes one specific obstacle that stood in the way of researchers who already wanted to do better and were defeated by the cost. Realigning what the research enterprise rewards is a larger project, and the collective statements calling for it are right to press it; this essay is not that project.

Some important concerns remain untouched by the argument: the provenance of the data these models were trained on, the attribution owed to the authors of that data, the widening gap between researchers who can afford frontier tools and those who cannot, and the temptation for institutions to read ``AI assistance'' as license to disinvest in people. These are serious, but not addressed here. They belong to a wider reckoning that the research-software community is right to demand. I mention them here so that the optimism of the preceding sections is read as bounded, not blind.

None of the conceded limits is a reason to wait: the skill-erosion risk is a reason to keep ownership, the non-determinism a reason to lean harder on tests, and the incentive problem a reason to advocate for better evaluation. Evidently, AI agents do not make research reproducible on their own. But the practices which make research reproducible now cost less and pay off immediately. That claim is narrow yet actionable.

\section{Agents on the Road to Reproducibility}

If you accept that agent-readiness and reproducibility work together, a corollary follows: whether your project gets real value from agentic AI is inseparable from whether your code is in a state that anyone---human or agent---can work with comfortably. A messy codebase will spawn messy agentic sessions. Researchers who report that AI tools are not living up to the promise may be seeing the consequences of a codebase that resisted structure for years. The practices that would have made their code defensible to a reviewer are the same practices that would now make it tractable to an agent.

When I publicly pledged a list of reproducibility practices for my research group, I did so with an optimism that wore down over the years---not because keeping to it was hard, but because it became tiresome to walk that road less travelled with so few others. Working reproducibly remained stubbornly unrewarded. The era now beginning lowers the labor cost of honoring those pledges while multiplying the returns. Today I would add a few items the 2012 Reproducibility PI Manifesto could not have imagined: that every repository we are responsible for carries an instruction file we keep current, so that no collaborator and no agent begins in ignorance; that every function whose correctness carries a scientific claim is pinned by a test we have read and trust; that every consequential decision is recorded with the reasoning that produced it, not merely the choice that resulted; that the history of each project explains why and not only what; and that we learn to drive these tools so the record of the work writes itself, under our control, as the work proceeds. These are new norms to follow and new practices to teach our students.

One pledge underwrites all the others, one that the growing concerns in our community demand: that we remain the owners and the verifiers of everything an agent drafts in our name. The artifacts are cheap now; the judgment is not, and the judgment is the part that was always ours to grow and defend. An agent can write the test, but only we can decide if it tests the right thing. An agent can record the decision, but only we can vouch that the reasoning is sound. That responsibility does not lighten when we delegate to agents; it actually grows, because there is more output to stand behind and less labor to remind us that we are the ones standing behind it.

For most of my career I have argued that researchers should adopt the practices of professional software engineering because doing so makes our science more reliable. I now make the case more readily, for a reason I would not have predicted: the test you once wrote out of duty, and skipped under deadline, you now write because an agent drafts it in seconds and uses it to check its own work an hour later. The payoff I used to ask people to take on faith can now be shown to be immediate. That, not any argument I could have made before, is what may change how we work from now on.

\bibliographystyle{unsrtnat}
\bibliography{references}
\end{document}